# Sentiment Polarity Detection for Software Development

Fabio Calefato[a], Filippo Lanubile[b], Federico Maiorano, Nicole Novielli[b]
[a] *Dipartimento Jonico, University of Bari "A. Moro", via Duomo, 259 -74123 Taranto, Italy.*
[b] *Dipartimento di Infomatica, University of Bari "A. Moro", via E. Orabona, 4 - 70125, Bari, Italy*
Email: {fabio.calefato, filippo.lanubile, nicole.novielli}@uniba.it, f.maiorano2@studenti.uniba.it

*Abstract* — The role of sentiment analysis is increasingly emerging to study software developers' emotions by mining crowd-generated content within social software engineering tools. However, off-the-shelf sentiment analysis tools have been trained on non-technical domains and general-purpose social media, thus resulting in misclassifications of technical jargon and problem reports. Here, we present Senti4SD, a classifier specifically trained to support sentiment analysis in developers' communication channels. Senti4SD is trained and validated using a gold standard of Stack Overflow questions, answers, and comments manually annotated for sentiment polarity. It exploits a suite of both lexicon- and keyword-based features, as well as semantic features based on word embedding. With respect to a mainstream off-the-shelf tool, which we use as a baseline, Senti4SD reduces the misclassifications of neutral and positive posts as emotionally negative. To encourage replications, we release a lab package including the classifier, the word embedding space, and the gold standard with annotation guidelines.

1. Introduction

Sentiment analysis is the study of the subjectivity (neutral vs. emotionally loaded) and polarity (positive vs. negative) of a text (Pang and Lee 2008). It relies on sentiment lexicons, that is, large collections of words, each annotated with its own positive or negative orientation (i.e., prior polarity). The overall sentiment of a text is therefore computed upon the prior polarity of the contained words.

Recent studies suggest approaches for enhancing software development, maintenance, and evolution by applying sentiment analysis on Stack Overflow (Rahman et al. 2015), app reviews (Maalej et al. 2016), and tweets containing comments on software applications (Guzman et al. 2016). Further research on developers' emotions addresses the role of affect in social software engineering, by applying sentiment analysis to the content available in collaborative development environments such as GitHub (Guzman et al. 2014, Guzman and Bruegge 2013, Sinha et al.2016), Jira (Mäntylä et al. 2016, Ortu et al. 2015), and Stack Overflow (Calefato et al. 2015, Novielli et al. 2015).

With a notable few exceptions (Blaz and Becker 2016, Panichella et al. 2015), empirical software engineering studies have exploited off-the-shelf sentiment analysis tools that have been trained on non-software engineering documents, such as movie reviews (Socher et al. 2013), or posts crawled from general-purpose social media, such as Twitter and YouTube (Thelwall et al. 2012). Jongeling et al. (2017) show how the choice of the sentiment analysis tool may impact the conclusion validity of empirical studies by performing a benchmarking study on seven datasets, including discussions and comments from Stack Overflow and issue trackers. By comparing the predictions of widely used off-the-shelf sentiment analysis tools, they show that not only these tools do not agree with human annotation of developers' communication channels, but they also disagree among themselves.

Another challenge to address is the negative bias of existing sentiment analysis tools, that is the misclassification of neutral technical texts as emotionally negative. It is particularly the case of bug reports or problem descriptions (Blaz and Becker 2016, Novielli et al. 2015). Novielli et al. (2015) show how sentences like "*What is the best way to kill a critical process*" or "*I am missing a parenthesis but I don't know where*" are erroneously classified as negative because both 'to kill' and 'missing' hold a negative polarity in the SentiStrength lexicon. This evidence is consistent with the *meaning-is-use* assumption that the sense of an expression is fully determined by its context of use (Wittgenstein 1965).

In this paper, we address the problem of applying sentiment analysis to the software engineering discipline. We propose a sentiment analysis classifier, named Senti4SD, which exploits a suite of lexicon-based, keyword-based, and semantic features (Basile and Novielli 2015) for appropriately dealing with the domain-dependent use of a lexicon. The approach implemented by Senti4SD successfully addresses the problem of misclassifying neutral sentences as negative. We observe a 19% improvement in precision for the negative class and a 25% improvement in recall for the neutral class with respect to the baseline, represented by SentiStrength. The emotion polarity classifier is publicly available[1] and represents the first contribution of this paper. To train and test Senti4SD, we built a gold standard of 4,423 posts mined from Stack Overflow. As a second contribution of this study, we release our gold standard as well as the emotion annotation guidelines to be used in further studies on emotion awareness in software engineering. Consistently with the *meaning-is-use* assumption, we assume that the contextual polarity of a word can be correctly inferred by its use. Thus, in order to derive our semantic features, we represent word meaning based on distributional semantics. In particular, we exploit neural-network-based approaches to distributional semantics, also known as word embedding (Levy and Goldberg 2014). Specifically, we used word2vec (Mikolov et al. 2013) to build a Distributional Semantic Model (DSM) where words are represented as high-dimensional vectors. The DSM, which builds on a collection of over 20 million questions, answers, and comments from Stack Overflow, represents a valuable resource for software engineering researchers who intend to investigate the use of word embedding in text categorization tasks. Therefore, we release the DSM as a third contribution of this study. Finally, as a fourth contribution, we provide a better understanding of the negative bias in off-the-shelf sentiment analysis tools when applied in the software engineering domain. The contribution of lexicon-based, keyword-based, and semantic features is confirmed by our empirical evaluation leveraging different feature settings. We provide empirical evidence of better performance also in presence of a minimal set of training documents.

The paper is structured as follows. In Section 2, we present an overview of the research methods followed by the theoretical background in Section 3. Section 4 describes the annotation study for building the gold standard. In Sections 5 and 6, we describe respectively the features used by our classifier, and then, the experimental setup and evaluation. Discussion and threats to validity are presented in Sections 7 and 8, respectively. In Section 9, we position our contribution with respect to related work. Finally, in Section 10 we draw conclusions and present future work.

## 2. Research Methods

Our research leverages a mix of qualitative and quantitative methods, including manual coding of textual data for building a gold standard on emotion polarity in software development, natural language processing techniques for feature extraction from Stack Overflow texts, and machine learning for training our emotion polarity classifier. Fig. 1 summarizes the process we followed in the current study. The complete process is organized in four sequential phases.

---

[1] The full lab package including Senti4SD, the DSM and the gold standard is available for download at: https://github.com/collab-uniba/Senti4SD

In Phase 1 we identified the theoretical framework of the current study and chose the emotion model to adopt in our annotation (see Section 3.1). The first output is the taxonomy of emotions and its mapping with polarity. As a second output, we defined the coding guidelines to adopt in the annotation study (see Appendix A).

In Phase 2 (see Section 4), the annotation study was carried out. We built the annotation sample by leveraging questions, answers, and comments extracted from Stack Overflow (see Section 4.1). Overall, the annotation sample is composed of 4,800 documents including questions, answers, and comments. User-contributed contents were preprocessed to improve their readability by discarding text elements that should not be annotated for sentiment, i.e. URLs, code snippets, and HTML tags. The annotation phase included the training of coders and a pilot annotation study before the final annotation was performed (see Sections 4.2 and 4.3).

In Phase 3 (see Section 4.3), we used the results of the annotation phase to build our gold standard for emotion polarity in software development. The interrater agreement was computed using Kappa, to assess the reliability of the annotation procedure and schema. The gold labels were assigned to documents in the annotation sample built using a majority voting criterion.

In Phase 4 (see Sections 5 and 6), we used the gold standard for emotion polarity to train and evaluate our classifier, whose performance was compared with off-the-shelf tools representing the state of the art for sentiment analysis on social media.

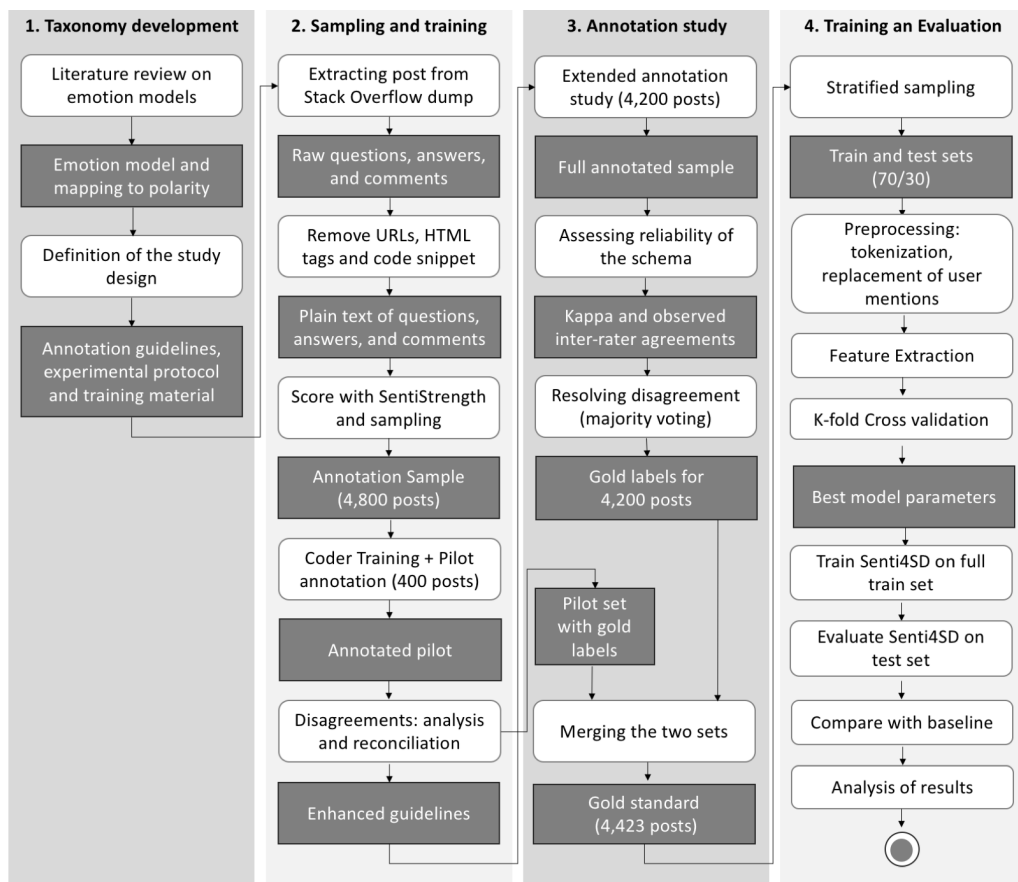

**Fig. 1** Overview of the research process

# 3. Background

In order to fully comprehend the addressed problem and the proposed solution, some key concepts are needed. The main points of such supporting concepts are presented in the following sections.

## 3.1. Emotion Modeling

Psychologists worked at decoding emotions for decades, developing theories based on cognitive psychology and natural language communication. So far, two points of view have emerged: one considers emotions as a continuous function of one or more dimensions, while the other assumes that a limited set of basic emotions exists (Carofiglio et al. 2009).

As regards the first viewpoint (continuous function), mining affective states from text typically involves modeling them across two dimensions: (1) the affect polarity, or valence, and (2) the level of activation, also known as arousal or intensity. It is the case of the 'circumplex model' of affect, which represents emotions according to a bi-dimensional representation schema capturing the emotion valence (pleasant vs. unpleasant) and arousal (activation vs. deactivation). According to this model, each emotion can be considered a "label for a fuzzy set, defined as a class without sharp boundaries" (Russell 1980).

On the other hand, theories following the discrete viewpoint agree on the idea that a limited set of basic emotions exists, although there is no consensus about the nature and the number of these basic emotions. According to Ekman (1999), basic emotions have specific feelings, universal signals, and corresponding physiological changes. Lazarus (1991) describes nine negative (anger, fright, anxiety, guilt, shame, sadness, envy, jealousy, and disgust) and seven positive (happiness, pride, relief, love, hope, compassion, and gratitude) emotions, with their appraisal patterns: positive emotions are triggered if the situation experienced is congruent with an individual goal, otherwise negative emotions are prompted.

Shaver et al. (1987) defined a tree-structured hierarchical classification of emotions. The hierarchy organizes emotion labels in three levels of hierarchical clusters. Each level refines the granularity of the previous one, thus providing more indication on its nature. The framework includes, at the top level, six basic emotions, namely love, joy, anger, sadness, fear, and surprise. The framework is easy to understand, thanks to the intuitive nature of the emotion labels. Consistently with our goal of training a classifier for emotion polarity, we map the emotions in the model by Shaver et al. to positive, negative, and neutral polarity (see Section 4). We use this mapping as a theoretical framework to inform our annotation guidelines (see Appendix A).

## 3.2. Polarity Detection with SentiStrength

SentiStrength (Thelwall et al. 2012) is a state-of-the-art, lexicon-based classifier that exploits a sentiment lexicon built by combining entries from different linguistic resources. In the SentiStrength lexicon, each negative word receives a sentiment score ranging from -2 to -5, which represents its prior polarity (i.e., the polarity of the term out of its contextual use). Similarly, positive words are associated with a score between +2 and +5, while neutral words receive scores equal to ±1. Positive and negative emoticons are also included in the dictionary. Based on the assumption that a sentence can convey mixed sentiment, SentiStrength outputs both positive and negative sentiment scores for any input text written in English. It determines the overall positive and negative scores to a text by considering the maximum among all the sentence scores, based on the prior polarity of their terms. Intensifiers, i.e., exclamation marks or verbs such as 'really', are treated as booster words and increase the word sentiment scores. Negations are also treated and determine the inversion of the polarity score for a given word. Therefore, the overall positive *p* and negative *n* sentiment scores issued by the tool range from ±1 (absence of positive/negative sentiment) to

±5 (extremely positive/negative). Based on their algebraic sum, SentiStrength can also report the overall trinary score, i.e. the overall positive (score = 1), negative (score = -1) and neutral (score = 0). Examples are provided in TABLE 1. The rationale for classification reported in the second column of the table is obtained by enabling the 'explain' option in SentiStrength.

TABLE 1. EXAMPLES OF SENTIMENT DETECTION IN STACK OVERFLOW USING SENTISTRENGTH.

| Input Text | Classification Rationale based on Word and Sentence Scores | Final Sentiment Score (maximum) | | Overall Score |
| --- | --- | --- | --- | --- |
| | | *Negative Score* | *Positive Score* | |
| "*I have very simple and stupid trouble*" | I have very simple and stupid [-3] trouble [-2] [sentence: 1, -3] [result: max + and – of any sentence] [overall result = -1 as pos<-neg] | -3 | 1 (absence of positive sentiment) | Negative (overall result = -1) |
| "*Thank you, that was really helpful*" | Thank [2] you, that was really helpful [2] [+1 booster word] [sentence: 3, -1] [result: max + and – of any sentence] [overall result = 1 as pos>-neg] | -1 (absence of negative sentiment) | +3 | Positive (overall result = 1) |
| "*I want them to resize based on the length of the data they're showing.*" | I want them to resize based on the length of the data they're showing. [sentence: 1, -1] [result: max + and - of any sentence] [overall result = 0 as pos=1 neg=-1] | -1 (absence of negative sentiment) | 1 (absence of positive sentiment) | Neutral (overall result = 0) |

Validated on social media, SentiStrength can deal with short informal texts that include abbreviations, intensifiers, and emoticons that typically occur in online interactions. As such, it has been widely adopted in social computing (Kucuktunc et al. 2012, Thelwall et al. 2012) and social software engineering (Guzman and Bruegge 2013, Guzman et al. 2014, Maalej et al. 2016, Novielli et al 2015).

To overcome the limitations and threats to validity derived from the use of off-the-shelf sentiment analysis tools in empirical software engineering studies (Blaz and Becker 2016, Jongeling et al. 2017, Novielli et al 2015), we train an emotion polarity classifier in a supervised machine learning setting by leveraging a gold standard of technical texts contributed by developers in Stack Overflow.

A customized version of SentiStrength has been developed to support sentiment analysis in software engineering (Islam and Zibran 2017). The tool is called SentiStrength-SE and is built upon the SentiStrength API. It leverages a manually adjusted version of the SentiStrength lexicon and implements *ad hoc* heuristics to correct the misclassifications observed when running SentiStrength on the Ortu dataset (Ortu et al. 2016). In our evaluation, we also include the performance of SentiStrength-SE for benchmarking (see Section 6).

### 3.3. Distributional Semantic Models

State-of-the-art sentiment analysis tools and lexicons rely on a dictionary-based word representation (see Novielli et al. 2015 for an overview). Words are treated as atomic units and are associated to a prior polarity expressed as a sentiment score, ranging from extremely negative to extremely positive with the absence of sentiment in the middle. Since the notion of word similarity is not taken into account, the polarity of a text is only based on the prior polarity of the words it contains and cannot be adjusted based on their contextual meaning.

Distributional Semantic Models (DSMs) represent words as mathematical points in high-dimensional vector spaces. A DSM relies on the so-called *distributional hypothesis* claiming that linguistic items with similar meanings occur in the same context

(Miller and Charles 1991). Based on the assumption that the meaning of a document is determined by the meaning of the words that appear in it, a text unit (e.g., a document, a sentence, a text fragment, etc.) can be represented as the vector sum of all the word vectors occurring in it. Thus, in a DSM, both words and documents are homogeneously represented as vectors and can be compared using similarity metrics that measure their closeness in the space, traditionally through cosine similarity (Mikolov et al. 2013b).

Traditional approaches to distributional semantics create word vectors by counting the occurrences of terms in a corpus and then operating a dimensionality reduction of word-by-document matrices. It is the case, for example, of Latent Semantic Analysis (Landauer and Dutnais 1997), which operates a singular value decomposition on the original term-by-document matrix to a low-dimension latent vector space. Such methods are usually referred in the literature as *context-counting* approaches (Baroni et al. 2014).

Recently, neural network-based approaches have been proposed (Bengio et al. 2003, Collobert and Weston 2008, Mikolov et al. 2013a) for learning distributed representation of words as continuous vectors. These approaches, also known as *word embedding* (Levy and Goldberg 2014), learn the vectors that maximize the probability of the contexts in which the target word appears. For this reason, they are usually referred to as *context-predicting* approaches (Baroni et al. 2014).

In our study, we leverage the approach defined by Milokov et al. (2013). They developed two models for implementing context-predicting approaches: (1) the Continuous Bag-of-Words (CBOW) model predicts the target word by considering the previous and following *n* words in a symmetrical context window; (2) the Skip-gram model predicts the surrounding words based on the target word. Both architectures are implemented in word2vec,[2] a publicly available tool for building a DSM from a large collection of documents. Both CBOW and Skip-gram models are capable of scaling up to large data sets with billions of words and are computationally more efficient for training high-dimensional spaces than context-counting approaches (Mikolov et al. 2013a). Furthermore, they outperform traditional context-counting approaches on standard lexical semantics benchmarks (Baroni et al. 2014).

## 4. Dataset: A Gold Standard for Emotion Polarity in Software Development

To train and evaluate our classifier for emotion polarity we built a gold standard composed of 4,423 posts from Stack Overflow. The dataset is well-balanced: 35% of posts convey *positive* emotions while 27% present *negative* emotions. No emotions are observed for the remaining 38% of posts, thus they receive the *neutral* polarity label.

In the following, we describe the sampling and coding processes adopted for building the gold standard.

### 4.1. Creating the Annotation Sample

The annotation sample was extracted from the official Stack Overflow dump of user-contributed content from July 2008 to September 2015. To improve their readability, we pre-processed all the posts, using regular expressions, to discard all those elements that are out of the scope of the sentiment annotation task, e.g. code snippets, URLs, and HTML tags.

In a previous study (Novielli et al. 2015), we found that stronger expressions of emotions are usually detected in comments rather than in question or answers. Therefore, we consider as a unit of analysis the Stack Overflow *post*, which includes not only questions and answers, but also comments provided by community members. Hence, conceptually we are addressing 3x4 groups of posts, that is, four types of Stack Overflow posts in `{question, answer, question comment, answer comment}` with three possible emotion styles in `{positive, negative, neutral}`.

---

[2] https://github.com/dav/word2vec

A desirable property of a training set is that its items are equally distributed across the existing classes of values (He and Garcia, 2009). Therefore, we built the dataset for the annotation by performing opportunistic sampling of posts based on both the presence of affectively-loaded lexicon and their type. To do so, we used SentiStrength to assess the presence/absence of affective lexicon in a post. We computed the positive and negative sentiment scores for the text of all the four types of posts extracted from the StackOverflow dump. Then, we randomly selected the same number of items based on the type of post and its sentiment scores. Our sample for annotation contains 4,800 items overall, equally distributed with respect to the types of posts and polarity, i.e. one-third of posts scored as positive by SentiStrength, one-third as negative, and one-third as neutral.

### 4.2. Pilot Annotation Study

Twelve coders participated in the emotion polarity annotation task. The coders were recruited among graduate CS students at the University of Bari and trained in a joint 2-hour session by the last author. She first explained the coding guidelines and then provided a sample of 25 Stack Overflow posts to be annotated individually in 30 minutes. Then, a follow-up discussion aimed at clarifying possible ambiguities in the interpretation of the coding guidelines.

Training was completed with a pilot subset of 100 items to be annotated individually at home. The twelve participants were organized into four groups of three coders each. Therefore, the pilot study was performed on 400 posts overall and each item in the dataset was assigned to three coders.

The coders were requested to indicate the emotion polarity, with a possible value in `{positive, negative, neutral, mixed}` (see Appendix A for guidelines). Analogously to Murgia et al. (2014), we refer to the Shaver et al. tree-structured framework for detecting emotions in the text (see Table A in Appendix A). The main difference with their annotation study is that we explicitly requested our coders to provide a polarity label, according to the specific emotion detected. In our study, *positive* polarity was indicated when the coders detected either joy or love. Conversely, *negative* polarity should be indicated when the coders identified anger, sadness, or fear. Regarding surprise, we asked the coders to determine the polarity based on contextual information. The *neutral* label indicates the absence of emotion. Posts conveying multiple emotions with opposite polarity (i.e., `joy` and `sadness`) were annotated as *mixed*.

The deadline was set a week after the assignment and the results of the annotation were discussed in a 2-hour plenary meeting with the experimenter. During this discussion, the coders had to resolve the disagreements on the pilot sample. This session was also used to disambiguate unclear parts of the guidelines as well as to enrich them with borderline examples whose annotation was agreed upon during the meeting. After the disagreements were solved, the pilot annotation became the first building block of the gold standard.

### 4.3. Emotion Polarity Coding: Extended Study

Once the training was complete, we assigned a new set of 500 posts to each coder. Once again, each item was annotated by three coders. Overall, 2,000 new items were annotated in this second step. Again, coders were required to perform this new annotation task individually. The deadline for returning the annotation was set in three weeks. We then assigned the final set of 600 posts to the coders. Overall, 2,400 additional new items were annotated in this final step.

As an evidence of the reliability of the coding schema and procedure, we computed the weighted Cohen's *Kappa* among the pairs of raters (Cohen 1968). We are interested in distinguishing between *mild* disagreement, that is the disagreement between negative/positive and neutral annotations, and *strong* disagreement, that is the disagreement between positive and negative judgments. We assigned a weight = 2 to strong disagreement and a weight = 1 to mild disagreement. We compute the inter-coder reliability for the entire set, including the pilot set annotation. The agreement is computed for all the four groups of participants (A, B, C, D) and for all pair of coders (C1, C2, and C3) in each group (see TABLE 2). We note a substantial agreement with *Kappa* values ranging in [.66, .80] (average .74). This evidence is confirmed also by the values of the *observed agreement*, which is the percentage of cases on which the raters agree, ranging in [.73, .85] (average .79).

Consistently with previous research on emotion annotation (Blaz and Becker 2016, Murgia et al. 2014), we resolved the disagreements by applying a majority voting criterion. We excluded from the gold standard all the posts for which opposite polarity labels were provided, including mixed cases (3%), even in presence of majority agreement. The final gold standard resulted in 4,423 posts, representing 92% of 4,800 annotated items.

TABLE 2. WEIGHTED COHEN'S KAPPA AND OBSERVED AGREEMENT FOR ALL THE COUPLE OF CODERS IN THE POLARITY ANNOTATION STUDY

| Group | *Weighted Cohen's Kappa for pairs of coders* | | | *Observed Agreement for pairs of coders* | | |
| --- | --- | --- | --- | --- | --- | --- |
| | C1 and C2 | C1 and C3 | C2 and C3 | C1 and C2 | C1 and C3 | C2 and C3 |
| A | .66 | .76 | .68 | .73 | .82 | .76 |
| B | .74 | .72 | .74 | .79 | .76 | .79 |
| C | .77 | .77 | .77 | .83 | .80 | .85 |
| D | .76 | .76 | .76 | .80 | .78 | .81 |
| | Average Weighted Cohen's Kappa .74 | | | Average Observed Agreement .79 | | |

## 5. Emotion Polarity Classifier: Feature Description and System Setup

Previous research shows how combining generic and domain-specific resources improves the performance of sentiment analysis (Bollegala et al. 2013). Therefore, we exploit three different kinds of features based on: (1) generic sentiment lexicons, (2) keywords (i.e., n-grams extracted from our dataset), and (3) word representation in a distributional semantic model specifically trained on software engineering data.

### 5.1. Lexicon-based features

The first set of features exploits existing sentiment lexicons. The approach is totally independent of the lexicon chosen and simply requires that a sentiment score is provided for each entry of the input (Novielli and Basile, 2015). For example, in the lexicon used by SentiStrength, each negative word is associated with an *a priori* sentiment score in [-2, -5]. Similarly, positive words receive a score in [+2, +5]. A list of objective words is also provided, with scores equal to ±1.

For a given post we compute the lexicon-based features reported in TABLE 3. In particular, we compute the number of tokens with positive and negative prior polarity (`Pos_words` and `Neg_words`), the overall number of tokens with either positive or negative prior polarity (`Subj_words`), the score of the last emoticon (`Last_emo`), the sum of all the scores for positive (`Sum_pos`), negative (`Sum_neg`), and subjective (`Sum_subj`) tokens, and the maximum positive and negative scores observed in the post (`Max_pos` and `Max_neg`). We also capture the presence of positive/negative utterances in combination with exclamation marks, indicating emphasis (`Pos_Emph` and `Neg_Emph`). Finally, we capture the sentiment of the last token/emotion (`End_Pos`, `End_Neg`), and whether it is combined with an exclamation mark (`End_Pos_Emph`, `End_Neg_Emph`). All the lexicon-based features have been already used for the sentiment analysis of crowd-generated content (Mohammad et al. 2013). Our lexicon-based features are independent of the specific lexicon adopted. To enable fair comparison with the baseline, represented by SentiStrength, for this study, we use the SentiStrength lexicon. The choice of the SentiStrength lexicon is further supported by its ability to deal short informal text, as it includes also abbreviations, intensifiers, and emoticons that are typically used in the social web. Furthermore, it incorporates sentiment scores from other linguistic resources that were previously validated in the scope of empirical research in sentiment analysis (Stone et al., 1966) and psycholinguistics (Pennebaker and Francis, 2001).

TABLE 3. SENTIMENT LEXICON-BASED FEATURES

| Feature | Description |
| --- | --- |
| Pos_words | The number of tokens with positive prior polarity. |
| Neg_words | The number of tokens with negative prior polarity. |
| Subj_words | The number of tokens with either negative or positive prior polarity. |
| Last_pos | The score of the last positive token in the post. |
| Last_neg | The score of the last negative token in the post. |
| Last_emo | The score of the last emoticon in the post. |
| Sum_pos | The sum of the score for the tokens with positive prior polarity. |
| Sum_neg | The sum of the score for the tokens with negative prior polarity. |
| Sum_subj | The sum of the score for the tokens with either positive or negative prior polarity. |
| Max_pos | The maximum score for the tokens with positive prior polarity in the post. |
| Max_neg | The maximum score for the tokens with negative prior polarity in the post. |
| Pos_emo | The number of emoticons with positive prior polarity. |
| Neg_emo | The number of emoticons with negative prior polarity. |
| Pos_Emph | Boolean, is true if the document has at least one positive token and ends with an exclamation mark, indicating emphasis. |
| Neg_Emph | Boolean, is true if the document has at least one negative token and ends with an exclamation mark, indicating emphasis. |
| End_Pos_Emph | Boolean, is true if the document ends with a positive token and an exclamation mark |
| End_Pos_Emph | Boolean, is true if the document ends with a negative token and an exclamation mark |
| End_Pos | Boolean, is true if the document ends with a positive token or emoticon. |
| End_Neg | Boolean, is true if the document ends with a negative token or emoticon. |

### 5.2. Keyword-based features

Keyword-based features include word counts for n-grams appearing in a document, in our case a Stack Overflow post. In our feature set, we consider uni- and bi-grams. Consistently with traditional approaches to text classification (Joachims 1998), each n-gram in our corpus corresponds to a feature, with the number of occurrences as its value. Other than including n-grams, we designed features able to capture aspects of micro-blogging, such as the use of uppercase and elongated words used as intensifiers, the presence of positive and negative emoticons, and the occurrence of slang expression of laughter (Basile and Novielli 2015). The total number of keyword-based features is 76,346. We report a summary in TABLE 4.

TABLE 4. KEYWORD-BASED FEATURES

| Feature | Description |
|---|---|
| *Uni-grams* | Total occurrences of uni-grams. Overall, our unigram dictionary counts 10,496 entries. |
| *Bi-grams* | Total occurrences of bi-grams. Overall, our bi-gram dictionary counts 65,844 entries. |
| *Uppercase_words* | Total occurrences of uppercase words (e.g. 'GOOD', 'BAD'). |
| *Laughter* | Total occurrences of slang expressions for laughter, such as 'hahaha' or abbreviations as 'LOL' occurring in the SentiStrength list of abbreviations. |
| *Elongated_words* | Total count of tokens with repeated characters (e.g. 'scaaaaary', 'gooooood'). |
| *M_repetitions* | The total occurrences of strings with repeated question or exclamation marks. (e.g., '????', '!!!!', '?!?!?!?!?'). |
| *User_mentions* | Total occurrences of user mentions (in the form @username). |
| *EndWith_EXMark* | Boolean, true if the document ends with an exclamation mark. |

### 5.3. Semantic features

The semantic features capture the similarity between the vector representations of the Stack Overflow documents and *prototype vectors* representing the polarity classes in a DSM. Analogously to (Basile and Novielli 2015, Novielli and Strapparava 2013), we represent a Stack Overflow document (i.e., a question, answer, or comment) in the DSM as the vector sum of all the vectors of words occurring in the document, using the superposition operator (Smolensky 1990).

The *prototype vectors* are vector representation of the positive, negative, and neutral classes in the DSM, namely `p_pos`, `p_neg`, and `p_neu`. A prototype vector is a vector representation of the lexical profile for a given polarity class, based on a sentiment lexicon that provides prior polarity scores for words. To compute the prototype vector `p_pos` for the positive class we sum all the vectors for words with positive polarity score in the chosen sentiment lexicon. In a similar fashion, we compute `p_neg` and `p_neu` by summing up, respectively, all the negative and neutral words in the chosen sentiment lexicon. In this study, we used the list of positive, negative, and neutral words included in the SentiStrength lexicon. We further calculated the subjective prototype `p_subj` vector by summing up the positive and negative word vectors, to better capture the differences in the lexical choice of neutral sentences and affectively-loaded ones.

We used the four prototype vectors to compute the semantic features, that is the similarity scores between the document vector (i.e., a Stack Overflow post) and each prototype vector, namely `Sim_pos`, `Sim_neg`, `Sim_neu`, `Sim_subj` (see TABLE 5). The semantic features are computed on a DSM built on Stack Overflow data, using the CBOW architecture implemented by word2vec (Mikolov et al. 2013), as depicted in Fig. 2. We choose a configuration with 600 vector dimensions, after having repeated the 10-fold cross-validation for parameter tuning (see Section 5.4.). We ran word2vec on a corpus extracted from the Stack Overflow official dump updated to September 2015. We extracted 3.8 million questions from the dump with the associated 5.9 million answers and 11.6 million comments. We preprocessed the posts to remove the URLs, HTML codes, and code snippets, obtaining a collection of more than 20 million posts with 912,201,785 tokens overall.

TABLE 5. SEMANTIC FEATURES

| Feature | Description |
|---|---|
| *Sim_pos* | The cosine similarity between the post vector and the objective prototype vector p_pos. |
| *Sim_neg* | The cosine similarity between the post vector and the objective prototype vector p_neg. |
| *Sim_neu* | The cosine similarity between the post vector and the objective prototype vector p_neu. |
| *Sim_subj* | The cosine similarity between the post vector and the subjective prototype vector p_subj. |

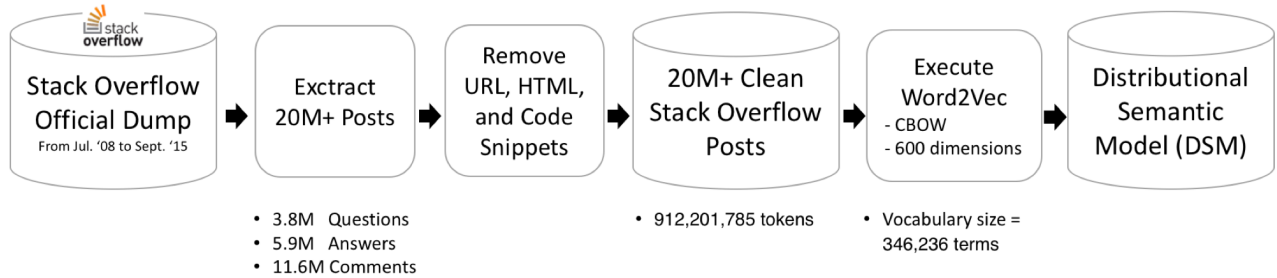

**Fig. 2** Building the DSM on Stack Overflow data

*5.4. System Setup and Parameter Tuning*

Before extracting all features, we performed tokenization using the Stanford NLP suite (Manning et al. 2014). During the tokenization, we replaced the user mentions with the meta-token `@USER`. We did not perform any stemming nor lemmatization since an inflected form may convey important information about polarity. It is the case, for example of 'ail' and 'ailing', which holds different prior polarity in the SentiStrength lexicon. Also, we did not remove stop words, consistently with previous research in sentiment classification tasks (Saif et al. 2014).

We trained Senti4SD using Support Vector Machines (SVM). SVM is able to learn and generalize even with a high dimensional feature space, which is a typical scenario in text classification tasks like ours (Joachims 1998). In particular, linear SVM is a state-of-the-art learning technique for such high-dimensional sparse datasets with a large number of items and a large number of features `N`, where each item has only `s << N` non-null features (Joachims 2006), as typical in presence of n-grams. One way to avoid dealing with such high dimensional input spaces would be to perform substantial feature selection. However, in supervised learning for text classification tasks, very few features are actually irrelevant, and feature selection results in a significant loss of information (Joachims 1998). Thus, we exploit the full set of features. Still, in order to assess the predictive value of our features, we analyze and rank them according to their information gain (Mitchell 1997). In TABLE 6, we report the top 25 features ranked by information gain.

We ran our experiment using the R interface (Helleputte 2015) to Liblinear (Fan et al. 2008), an open source library for large-scale linear classification with SVM. For linear classification, the only parameter is the cost parameter C. Too large C values makes the cost of misclassification high, thus forcing the algorithm to better explain the training data but potentially inducing the risk of overfitting. To fine-tune the SVM parameter while still preventing overfitting, we ran the Liblinear parameter tuning utility on our training set in a 10-fold cross-validation setting. We chose the optimal value for the C parameter by maximizing the prediction accuracy. We repeated the parameter tuning with all the settings derived by combining the two available architectures in word2vec CBOW and Skip-gram, with vector space dimensions in `{200, 400, 600, 800, 1000}`. Furthermore, word2vec has two input parameters for rare-word pruning and frequent word sub-sampling: words appearing less than `min-count` times are discarded from the document collection before starting the DSM training, while frequent words (as defined by the `sample` input parameter) are down-sampled to increase the effective context window considered for vector prediction (Mikolov et al. 2013). Consistently with previous research (Basile and Novielli 2015), we maintained the default value for the `sample` parameter while discarding the terms with less than 10 occurrences, thus obtaining a final vocabulary of 346,236 terms. The optimal configuration, used to train our final classifier over the training set, is reported in TABLE 7.

TABLE 6. TOP 25 FEATURES RANKED BY INFORMATION GAIN

| Rank | Feature | Information Gain | Feature Group |
|---|---|---|---|
| 1 | Sum_pos | 0.56081 | Lexicon-based |
| 2 | Max_pos | 0.54642 | |
| 3 | Pos_words | 0.52497 | |
| 4 | Last_pos | 0.51273 | |
| 5 | Sum_neg | 0.39355 | |
| 6 | Max_neg | 0.3933 | |
| 7 | Neg_words | 0.3876 | |
| 8 | Last_neg | 0.38291 | |
| 9 | Sum_subj | 0.33765 | |
| 10 | Subj_words | 0.31614 | |
| 11 | Sim_pos | 0.26473 | Semantic |
| 12 | Sim_neg | 0.16507 | |
| 13 | Sim_subj | 0.15596 | |
| 14 | Sim_obj | 0.11297 | |
| 15 | Last_emo | 0.10775 | Lexicon-based |
| 16 | Pos_emo | 0.07299 | |
| 17 | 'great' | 0.06496 | Keyword-based |
| 18 | 'excellent' | 0.06055 | |
| 19 | ':)' | 0.05649 | |
| 20 | 'thanks' | 0.03856 | |
| 21 | Neg_emo | 0.03525 | Lexicon-based |
| 22 | 'hate' | 0.03445 | Keyword-based |
| 23 | 'annoying' | 0.0282 | |
| 24 | Uppercase_words | 0.02819 | |
| 25 | ':(' | 0.028 | |

TABLE 7. SYSTEM SETUP AFTER PARAMETER OPTIMIZATION

| | Parameter | Value |
|---|---|---|
| DSM | word2vec architecture | Continuous Bag-of-Words (CBOW) |
| | DSM dimensions | 600 |
| SVM | C | 0.05 |

## 6. Evaluation

### 6.1. Creation of Train and Test Sets

We split the gold set into training (70%) and test (30%) sets, using the R (R Development Core Team 2008) package *caret* (Kuhn 2016) for stratified sampling. We used the training set to seek the optimal parameter setting for our classifier (see Section 5.4). The final model was trained on the whole training set using the optimal configuration and then evaluated on the test set, to assess to what degree the trained model is able to generalize sentiment polarity classification on unseen new data from the held-out test set.

### 6.2. Results

After having trained Senti4SD, we evaluated the learned model on the Stack Overflow test set. TABLE 8 reports the performance obtained in terms of recall, precision, and F-measure for the single classes and overall. The overall performance is computed adopting micro-averaging as aggregated metric (Sebastiani 2002). We highlight in bold the best value for each metric.

In TABLE 8, we also report the performance of SentiStrength[3] on the Stack Overflow test set, which we consider as a baseline for the performance assessment of Senti4SD. We choose SentiStrength because it is the most widely employed tool in sentiment analysis studies in software engineering (Calefato et al. 2015, Guzman et al. 2016, Guzman and Bruegge 2013, Ortu et al. 2015, Sinha et al. 2016). In addition, we also report the performance of SentiStrength-SE. We mapped both SentiStrength and SentiStrength-SE scores to a categorical sentiment label in {positive, neutral, negative} for each entire question, answer or comment. Consistently with the approach defined by SentiStrength authors (Thelwall et al. 2012) and already adopted in previous benchmarking studies (Jongeling et al. 2017), given the positive (p) and negative (n) scores issued by the tool, we consider a text as `positive` when `p + n > 0,` `negative` when `p + n <0,` and `neutral` if `(p = n) and (p < 4)`. Texts with a score of `p = n` and `p ≥ 4` are considered having an undetermined sentiment and should be removed from the dataset. However, no such controversial cases are found in our dataset. Looking at the performance of Senti4SD, we observe a 19% improvement in precision for the negative class and a 25% improvement in recall for the neutral class with respect to the SentiStrength baseline, which in turn outperforms SentiStrength-SE.

TABLE 8. PERFORMANCE OF SENTI4SD AND COMPARISON WITH SENTISTRENGTH (BASELINE).

|  | *Overall* | | | *Positive* | | | *Negative* | | | *Neutral* | | |
| --- | --- | --- | --- | --- | --- | --- | --- | --- | --- | --- | --- | --- |
|  | *R* | *P* | *F* | *R* | *P* | *F* | *R* | *P* | *F* | *R* | *P* | *F* |
| Baseline (SentiStrength) | .82 | .82 | .82 | **.92** | .89 | .90 | **.96** | .67 | .79 | .64 | **.95** | .76 |
| SentiStrength-SE | .78 | .78 | .78 | .79 | .90 | .84 | .79 | .73 | .76 | .77 | .73 | .75 |
| Senti4SD | **.87** | **.87** | **.87** | **.92** | **.92** | **.92** | .89 | **.80** | **.84** | **.80** | .87 | **.83** |
| *Improvement over the SentiStrength baseline* | *+6%* | *+6%* | *+6%* | *---* | *+3%* | *+2%* | *-7%* | *+19%* | *+6%* | *+25%* | *-8%* | *+9%* |

TABLE 9. AGREEMENT BETWEEN MANUAL LABELING AND PREDICTION ON THE STACK OVERFLOW TEST SET.

|  |  | *Prediction* | | | | | |
| --- | --- | --- | --- | --- | --- | --- | --- |
|  |  | **SentiStrength** | | | **Senti4SD** | | |
|  |  | Negative | Positive | Neutral | Negative | Positive | Neutral |
| *Manual* | Negative | 345 (95.8%) | 7 (1.9%) | 8 (2.2%) | 321 (89.2%) | 3 (0.8%) | 36 (10.0%) |
|  | Positive | 30 (6.6%) | 420 (91.7%) | 8 (1.8%) | 11 (2.4%) | 423 (92.4%) | 24 (5.2%) |
|  | Neutral | 140 (27.6%) | 44 (8.7%) | 324 (63.8%) | 70 (13.8%) | 32 (6.3%) | 406 (79.9%) |

---

[3] The evaluations have been performed using the SentiStrength Java API obtained from http://sentistrength.wlv.ac.uk/ on December 2016.

In TABLE 9 we report the confusion matrix for both SentiStrength and Senti4D, showing the agreement between the manual labeling and the polarity predicted by each tool. We consider only SentiStrength because it outperforms SentiStregth-SE. We complement the evidence provided by the confusion matrix with Venn diagrams representing the posts correctly classified as negative, positive, and neutral by SentiStrength and Senti4SD (see Fig. 3). Looking at the predictions, we observe that the 24 negative cases recognized only by SentiStrength (see Fig. 3.a) are classified as neutral by Senti4SD. As for positive posts (see Fig.3.b), the 10 cases missed by Senti4SD are classified mostly as neutral and only one of them is classified as negative. Conversely, the 13 recognized only by Senti4SD are misclassified by SentiStrength as positive. As for neutral, the 84 cases recognized only by Senti4SD (see Fig.3.c) are classified mainly as negative (69/84).

We complement the previous evaluation with an assessment of the advantage of including all the features defined in Senti4SD. Our goal is to assess whether the improvement of performance, with respect to SentiStrength, is a result of the adopted machine learning technique or is rather due to the additional features (full set of keyword-based, semantic features, and lexicon-based features) we propose. We start by computing the performance with a simple model including only the uni- and bi-grams. Such a model does not include consideration of any sentiment-specific feature and represents the traditional approach to text categorization based on machine learning (Joachims 1998). By doing so, we want to assess to what extent the additional features contribute to the performance by capturing sentiment-related linguistic phenomena. Then, we evaluate the performance of Senti4SD by considering incremental feature settings, in order to assess the contribution of each feature group to the classifier performance. Results are reported in TABLE 10 and complement the evidence provided by the information gain analysis (see TABLE 6) about the role of each feature group. The last column of TABLE 10 (*p-value <0.05*) indicates whether we observe a statistically significant improvement in a given setting over the previous approach. Statistical significance of the difference between settings is computed by performing the Chi-squared test with α= 0.05.

By comparing the performance of Senti4SD leveraging different sets of features, we observe that simply training a supervised classifier based on n-grams does not yield an acceptable overall performance (F = .69). By leveraging the full set of keyboard-

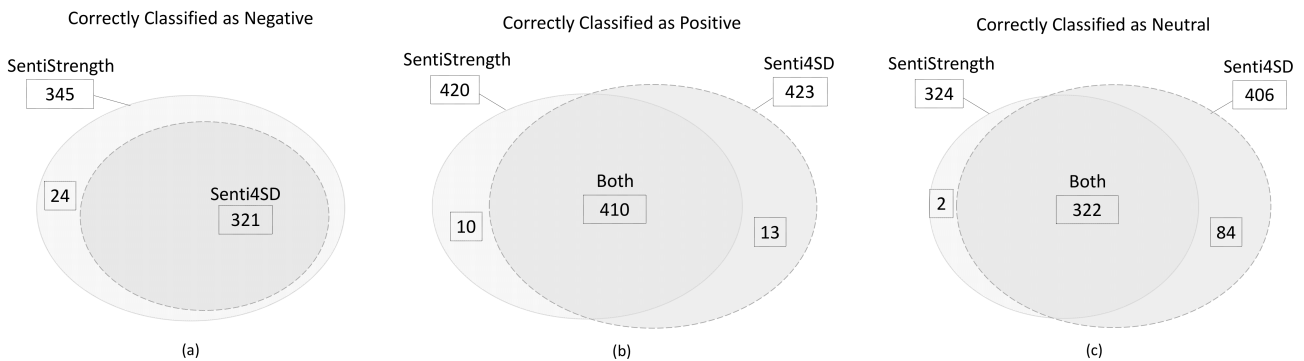

**Fig. 3** Posts Correctly Classified as Negative (a), Positive (b), and Neutral (c) by Senti4SD and SentiStrength

TABLE 10. PERFORMANCE OF SENTI4SD WITH INCREMENTAL FEATURE SETTINGS.

| Experimental Setting | Overall | | | Positive | | | Negative | | | Neutral | | | p-value <0.05 |
|---|---|---|---|---|---|---|---|---|---|---|---|---|---|
| | *R* | *P* | *F* | *R* | *P* | *F* | *R* | *P* | *F* | *R* | *P* | *F* | |
| N-grams only | .69 | .69 | .69 | .84 | .75 | .79 | .88 | .57 | .69 | .42 | .85 | .56 | |
| Keyword-based features | .79 | .79 | .79 | .84 | .84 | .84 | .67 | .80 | .73 | .83 | .74 | .79 | * |
| Keyword + Semantics | .81 | .81 | .81 | .86 | .86 | .86 | .73 | .80 | .76 | .83 | .78 | .81 | * |
| Keyword + Semantics + Lexicon Based (full feature set) | .87 | .87 | .87 | .92 | .92 | .92 | .89 | .80 | .84 | .80 | .87 | .83 | * |

* $p < 0.05$

based features the overall performance improves to F = .79. However, such a classifier would still perform poorly if compared to the Sentistrength baseline. In particular, recall for *negative* is unsatisfying (R = .67). Adding the four semantic features, performance is further improved (F=.81). In particular, adding only four features (i.e., the cosine similarity of the document with the sentiment prototype vectors) we observe an improvement of the *negative* recall from .67 to .73 and of the *neutral* precision from .74 to .78. However, while the improvement is statistically significant, the recall for *negative* is still low (R=.73). Finally, we include consideration of lexicon-based features, which further increase the recall of *negative* up to .89 and the precision of *neutral* up to .87.

Searching for further evidence of the robustness of the approach implemented by Senti4SD, we also assess its performance by splitting our gold standard into train and test set with different percentages. Results are reported in TABLE 11 and compared with the SentiStrength performance on the same test set for each iteration. For each setting, we compare the behavior of the two classifiers by performing a Chi-square test on the predictions issued by Senti4SD and SentiStrength, observing a p-value lower than 0.05. In all the three settings, Senti4SD outperforms SentiStrength, even with a reduced training set (last row of TABLE 11), with 30% of the gold standard used as training set. Again, we highlight in bold the best value for each metric.

## 7. Discussion

**Comparison with the SentiStrength baseline.** The performance of SentiStrength on our test set (see TABLES 8 and 9) confirms previous findings about its negative bias in the software engineering domain (Novielli et al. 2015). In the case of our

TABLE 11. PERFORMANCE OF SENTI4SD AND COMPARISON WITH SENTISTRENGTH WITH DIFFERENT TRAIN/TEST PROPORTIONS.

| Experimental Setting | Classifier | Overall | | | Positive | | | Negative | | | Neutral | | |
|---|---|---|---|---|---|---|---|---|---|---|---|---|---|
| | | *R* | *P* | *F* | *R* | *P* | *F* | *R* | *P* | *F* | *R* | *P* | *F* |
| Train = 70% Test = 30% (same as TABLE 8) | SentiStrength | .82 | .82 | .82 | .92 | .89 | .90 | **.96** | .67 | .79 | .64 | **.95** | .76 |
| | *Senti4SD* | *.87* | *.87* | *.87* | *.92* | *.92* | *.92* | *.89* | *.80* | *.84* | *.80* | *.87* | *.83* |
| Train = 50% Test = 50% | SentiStrength | .82 | .82 | .82 | **.93** | .89 | **.91** | .95 | .68 | .79 | .64 | **.94** | .77 |
| | *Senti4SD* | *.85* | *.85* | *.85* | *.91* | *.89* | *.90* | *.87* | *.80* | *.83* | *.78* | *.84* | *.81* |
| Train = 30% Test = 70% | SentiStrength | .82 | .82 | .82 | **.93** | .90 | .91 | **.94** | .67 | .78 | .64 | **.93** | .76 |
| | *Senti4SD* | *.84* | *.84* | *.84* | *.91* | *.91* | *.91* | *.83* | *.78* | *.80* | *.79* | *.82* | *.80* |

Stack Overflow dataset, SentiStrength erroneously classifies 28% of neutral posts as negative, with a poor recall for neutral class (.64) and a low precision for the negative one (.67). Since Stack Overflow is explicitly designed to support developers looking for help, discussions are often misclassified as conveying negative polarity because they are naturally rich in the 'problem' lexicon, which does not necessarily indicate the intention to show any affective state.

Senti4SD is able to address the problem of such negative bias. TABLE 9 shows that the number of neutral documents misclassified as negative is reduced from 27.6% in SentiStrength to 13.8% in Senti4SD. As a consequence, the F-measure increases from .79 to .84 for the negative class and from .76 to .83 for the neutral class, thus depicting a more balanced classifier (see TABLE 8). In particular, our classifier improves the recall of neutral documents from .64 up to .80 (25% of improvement) and the precision of negative documents from .67 up to .80 (19% of improvement). For example, SentiStrength erroneously classifies as negative sentences that are instead neutral, as the following ones: '*This will help you to come back to the previous activity. As per your code, the application was completely killed.*', '*Or if you don't want to worry about height calculation do this*'. On the contrary, Senti4SD correctly labels the above sentences as neutral.

However, we observe that this gain in precision is obtained at the expense of the negative class recall, which decreases from .96 to .89. For example, SentiStrength correctly classifies as negative the following posts: '*Is it possible to prevent a user from editing the title of a node on the node edit screen? One of the things I really detest about Drupal is the rigidity of the title & body field in each node*' and '*Ew, that sounds a bit ugly! Is it possible for an instance of a class to be created before its unit's initialization section has run? In other words, could an instance of TMyObject try to use FLogger before it's been set in the initialization section?*'. On the contrary, the two posts are erroneously labeled as neutral by Senti4SD even in presence of the negative lexicon (i.e., '*detest*' and '*ugly*'). Such misclassifications are probably due to the prevalence of neutral lexicon in the posts. Specifically, in the first post the first sentence does not carry any sentiment while in the second post the second and third sentences are neutral. A possible way to overcome this limitation occurring with long posts is to perform finer-grained annotation at a sentence level in order to train a sentence-based version of Senti4SD.

Misclassification of positive posts as negative occurs in 6.6% of the cases when the classification is performed with SentiStrength (see TABLE 9). This is what we consider a strong disagreement that should be avoided. Senti4SD reduces such misclassification to 2.4% of the cases. For example, sentences like '*Is in so u need not worry! Internally the data is always stored as TEXT, so even if you create table with, SQLite is going to follow the rules of data type*' is erroneously classified by SentiStrength as negative due to the presence of negative lexicon ('worry') even if SentiStrength is supposed to correctly deal with negations (Thelwall et al. 2012) which should determine polarity inversion.

Surprisingly, SentiStrength-SE produces a lower performance than SentiStrength on our Stack Overflow gold standard albeit it outperformes SentiStrength on other technical texts (Islam and Zibran 2017). This might occur because SentiStrength-SE incorporates *ad hoc* heuristics and word polarity scores that are specifically designed to solve misclassifications observed on a small unbalanced dataset of 400 developers' comments in Jira (Ortu et al. 2016). As such, overfitting is a plausible explanation for the decay in performance of SentiStrength-SE in our study.

**Implications.** Senti4SD was developed in the scope of our ongoing research on the role of emotions in social software engineering (Novielli et al. 2014, Mäntylä et al. 2017). More specifically, we envision the emergence of sentiment analysis tools monitoring communication between the developers as well as user-contributed technical texts (e.g., reviews in app stores), analyzing the affect expressed in this communication and translating the results into actionable insights (Gachechiladze et al 2017). Among others, negative affective states deserve attention because of their detrimental impact on developers' productivity

(Denning 2012, Ford and Parnin, 2015, Graziotin et al. 2017). When implementing a sentiment classifier, deciding whether to optimize by precision or by recall is not a trivial decision, which depends on the application scenario. Early detection of negative sentiment towards self, such as frustration, could be useful to design tools for supporting developers experiencing cognitive difficulties (i.e., learning a new language or solving tasks with high reasoning complexity) (Ford and Parnin, 2015), as well as in their daily programming tasks (Müller and Fritz 2015). In such a scenario, a monitoring tool might suggest the intervention of an expert or provide a link to further material and documentation to support the developer. However, a sentiment analysis tool with high recall and low precision for negative sentiment as SentiStrength would produce several false positives, causing undesired, erroneous interruptions that are detrimental to developers' productivity and focus. In such cases, being able to reduce the number of false positives for negative sentiment becomes crucial and Senti4SD should be preferred to SentiStrength, due to its higher precision.

Similarly, timely detection of negative sentiment towards peers, such as anger and hostility (Gachechiladze et al 2017), might be exploited for detecting code of conduct violations (Tromp and Pechenizkiy 2015) or enhancing effective community management. For example, sentiment analysis may support GitHub users who want to be notified of heated conversation and lock them before flame wars break out.[4] In scenarios that involve human intervention to guide the contributors' behavior towards a constructive pattern, it might be desirable to optimize negative sentiment detection by recall, thus choosing to leverage SentiStrength higher sensitivity to negative emotions. Conversely, if automatic filtering of offensive comments or conversation is envisaged, it becomes important to optimize by precision by using Senti4SD, to avoid banning neutral conversations.

Finally, sentiment analysis is now regarded as a technique also useful for mining large software repositories, e.g., to understand the role of sentiment in security discussions (Pletea et al. 2014) and commits in GitHub (Guzman et al. 2014). In such scenarios, a sentiment classifier specifically trained and validated in the software engineering domain allows controlling for threats to validity due to inappropriate instrumentation, as argued by Jongeling et al (2017).

**Contribution of features.** Consistently with traditional approaches to supervised machine learning in text classification (Joachims 1998), we did not perform feature selection, thus including in our evaluation setting the full suite of lexicon-based, keyword-based, and semantic features described in Section 5. As a further evidence of the importance of each group of features, we performed an analysis based on information gain (see TABLE 6) and assessed Senti4SD performance by leveraging different feature settings (see TABLE 10). The top-ten predictive features belong to the group of lexicon-based, which is an expected result since they are based on sentiment lexicons specifically designed to represent the sentiment polarity association to words. They are immediately followed by the four semantic features that measure the similarity between a document and the linguistic profile of each polarity class. Among the top predicting keyword-based features, we find positive and negative emoticons (`Pos_emo` and `Neg_emo`, respectively). Expressions of gratitude (i.e., `'thanks'`) and appreciation (i.e., `'great'`, `'excellent'`) are also among the top uni-gram predictors, thus confirming evidence from previous research that paying gratitude for the help received as well as enthusiasm for the solution provided are the main causes for positive sentiment in the social programmer ecosystem (Calefato et al. 2015, Novielli et al. 2015, Ortu et al. 2015). Conversely, expression of anger and frustration (i.e., `'hate'`, `'annoying'`) are among the top predictors for negative sentiment. The contribution of each feature group to the classification performance is confirmed by the evaluation of Senti4SD leveraging different feature settings, as reported in TABLE 10. Furthermore, we provide empirical evidence that supervised training as implemented by Senti4SD produce similar performance also in presence of a minimal set of training documents (see TABLE 11).

---

[4] https://help.github.com/articles/locking-conversations

**Gold standard.** Our manually annotated dataset is the first resource on emotion polarity to be built upon the corpus of Stack Overflow. As such, our dataset represents a valuable resource in the scope of empirical research on emotion awareness in software engineering (SEmotion 2016). Stack Overflow is an example of an online community where programmers do networking by reading and answering others' questions, thus participating in the creation and diffusion of crowdsourced documentation. Among the non-technical factors that can influence the members of online communities, the emotional style of a technical contribution does affect its probability of success (Calefato et al. 2015). Being able to identify harsh comments towards technical matters could be useful in detecting particularly challenging questions that have not been exhaustively answered (Novielli et al. 2015), which is a goal addressed by current research on effective knowledge-sharing (Anderson et al. 2012). Similarly, detecting negative attitude towards the interlocutor could allow the community moderators to guide users towards appropriate interaction patterns. This is an open problem in the Stack Overflow community, as users complain about harsh comments coming from expert contributors (Meta 2017), which may impair successful question-answering (Asaduzzaman et al. 2013).

The release of our gold standard complements the effort of Ortu et al. (2016) who recently released a dataset of 2,000 issue comments and 4,000 sentences written by developers, collected by mining the repositories of four open source ecosystems, namely Apache, Spring, JBoss, and CodeHaus. Their dataset is annotated using the basic emotion labels in the framework by Shaver et al. (1987) that we also adopt in the present study.

## 8. Threats to Validity

Our methodology could produce different results if applied outside of Stack Overflow. However, Stack Overflow is so popular among software developers (currently used by about 7 million software developers[5]) to be reasonably confident that the dataset is representative of developers' communication style. Nevertheless, we acknowledge that replications are needed to further increase the external validity to the entire software developer ecosystem.

We built our gold standard on emotion polarity through manual annotation. Emotion annotation is a subjective process since affect triggering and perception can be influenced by personality traits and personal dispositions (Scherer et al. 2004). To mitigate this threat, we provided clear guidelines (see Appendix A) grounded on a theoretical framework for emotion identification based on the model by Shaver et al. (1987). Furthermore, polarity labels were assigned using majority agreement among three coders. To be more conservative, even in presence of majority agreement, we excluded from the gold standard all the posts for which opposite polarity labels were provided. The interrater agreement (average weighted Cohen's Kappa = 0.74) confirms a good reliability of the gold standard. Nevertheless, we intend to improve coding guidelines by enriching the number of examples, especially for those more controversial that lead to coding conflicts.

The sample set for the emotion annotation experiment was built using SentiStrength. We built our sample set to have one-third of posts scored by SentiStrength as positive, one-third as negative, and one-third as neutral. However, as highlighted by previous research (Blaz and Becker 2016, Jongeling et al. 2017, Novielli et al. 2015), off-the-shelf tools for sentiment analysis report limited performance when detecting sentiment in the software engineering domain. In particular, SentiStrength tends to misclassify neutral sentences as negative (see TABLES 8 and 9). As a result, we ended up including in our sample set a higher proportion of neutral sentences that were originally misclassified as negative by SentiStregth and later correctly classified as neutral by our coders. Another cause of error when using SentiStrength on our data is the misclassification of positive posts as negative (see TABLE 9). As such, a small proportion of the posts originally included in the sample annotation set because rated as negative by SentiStrength were subsequently classified as positive by the coders. The distribution of the gold standard built through the annotation study confirms these issues and shows how the negative class is underrepresented in our dataset, i.e., the gold standard contains 35% *positive* posts, 27% *negative* posts, and 38% of neutral posts. Another consequence of using SentiStrength to create the sample set is that the sentences in our dataset contain emotion words included in the SentiStrength lexicon. Hence, we observe a very good performance of the tool on our gold standard (F = .82, TABLE 8), making SentiStrength a challenging baseline for our classification task.

---
[5] Source: http://stackexchange.com/sites#questions Last accessed: June '17

Finally, we excluded from the gold standard all the posts for which opposite polarity labels were provided, which represent the 3% of all annotated data. Our choice is justified by the intention to not introduce noise in the data during the supervised training phase of Senti4SD. Currently, Senti4SD would classify those posts as either positive or negative. However, we acknowledge that a minority of posts might present both positive and negative emotions. In our future work, we will fine-tune Senti4SD by training separate binary classifiers for positive and negative sentiment to be able to recognize also mixed sentiment.

## 9. Related Work

### 9.1. Sentiment Analysis Resources for Software Engineering

Trying to overcome the limitations posed by using off-the-shelf sentiment analysis tools, software engineering researchers recently started to develop their own tools.

Panichella et al. (2015) applied sentiment analysis for classifying user reviews in Google Play and Apple Store. They trained their own classifier on 2,000 manually-annotated reviews, using Naïve Bayes and a bag-of-word approach. However, they do not report evaluation metrics for their classifier so we are not able to make any comparison with their method. For the sake of completeness, we also experimented with Naïve Bayes, as they suggest, but we found that it is outperformed by SVM.

Mäntylä et al. (2016) investigated the potential of mining developers' emotions in issue-tracking systems to prevent loss of productivity and burnout. They measured the emotions in issue comments in terms of VAD metrics, that is, scores for the Valence (i.e., the affect polarity), Arousal (i.e., the affect intensity), and Dominance (i.e., the sensation of being in control of a situation). To estimate VAD scores, they adopted the same lexicon-based approach implemented by SentiStrength, using a VAD lexicon of over 13K English words developed by psychology research. However, given the lack of a gold standard for VAD, they were not able to provide any evaluation of their approach to emotion mining.

Ortu et al. (2015) presented an empirical study on the correlation of emotions and issue-fixing time in the Apache issue-tracking system. They measure the emotion polarity in issue comments using SentiStrength. As for discrete emotion labels, they developed their own classifier for detecting the presence of four basic emotions framework by the Shaver et al. (1987), namely anger, joy, sadness, and love. Their approach exploits SVM using a suite of features based on the SentiStrength output, the politeness score (Danescu-Niculescu-Mizil et al. 2013), and the presence of affective words derived from WordNetAffect (Strapparava and Valitutti 2004). The classifier is evaluated on a gold standard of 4,000 sentences, obtaining an F-measure score ranging from .74 for anger to .82 for sadness. At the time of writing, the classifier is not yet available for research purposes.

Blaz and Becker (2016) developed a polarity classifier for IT tickets. Their approach is based on a domain dictionary created using a semiautomatic bootstrapping approach to expanding an initial set of affectively-loaded words used as seeds. They also exploit features based on the document structure, i.e., by distinguishing between the polarity of the opening section from the polarity of the actual problem report section in the ticket. They compare different approaches with different feature settings. In the best setting, they obtain an overall performance of F = .85, that is comparable to the one achieved by Senti4SD (F =.86). However, their classifier still reports a negative bias inducing the misclassification of neutral documents as negative. Their performance on the negative class (F = .70, R = .74, P = .67) reflects such bias. Senti4SD successfully addresses this problem by obtaining a more balanced performance for both the negative (F = .84, R = .87, P = .80) and neutral (F = .83, R = .80, P = .85) classes. Furthermore, Senti4SD has been trained and evaluated on a balanced and larger dataset.

Islam and Zibran (2017) developed SentiStrength-SE, a customized version of SentiStrength for software engineering. The tool is built upon the SentiStrength API and incorporates *ad hoc* heuristics designed to solve the misclassifications of SentiStrength observed on a subset of about 400 comments from the Ortu dataset (Ortu et al. 2016). The sentiment scores of the lexicon have been manually adjusted to reflect the semantics and neutral polarity of domain words such as '*support*', '*error*', or '*default*'. The authors performed an evaluation of the tool on the remaining 5,600 comments from the Ortu dataset, showing that SentiStrength-SE (F = .78, R = .85, P = .74) outperforms SentiStrength (F = .62, R = .79, P = .62) on technical texts. However, SentiStrength-SE produces a lower performance (F = .78, R = .78, P = .78) than both SentiStrength (F = .82, R = .82, P = .82) and Senti4SD (F = .87, R = .87, P = .87) when used to classify polarity of posts from our Stack Overflow gold standard (see Section 6).

### 9.2. Distributional Semantics in Software Engineering

To the best of our knowledge, word embedding techniques have not been applied before to sentiment analysis tasks in the software development domain. In particular, we exploit the idea of using features based on the document similarity with respect to prototype vectors modeling the lexical profile of the polarity classes. The use of prototype vectors in text classifications was successfully exploited in different domains, e.g., for unsupervised speech-act recognition in telephone conversations (Novielli and Strapparava 2013) and for sentiment analysis in micro-blogging (Basile and Novielli 2015).

However, the use of distributional semantics is not entirely new in software engineering research. Traditional, context-counting approaches to distributional semantics have already been used, including Latent Dirichlet Allocation for topic modeling in Stack Overflow (Barua et al. 2014) and Latent Semantic Analysis for recovering traceability links in software artifact (De Lucia et al. 2007). Tian et al. (2014) recently proposed the use of pointwise mutual information to represents word similarity in a high-dimensional space. They built SEWordSim, a word similarity database trained on Stack Overflow questions and answers. However, as already discussed in Section 2.2, count-based approaches suffer from the main drawback of poor scalability. In the specific case of SEWordSim, the words are represented in a high-dimensional matrix whose $e_{ij}$ elements correspond to the pointwise mutual information between the words $w_i$ and $w_j$, thus describing their semantic association. The fact that vector space dimensions equal the vocabulary size significantly limits the scalability of approaches based on such word models. Instead, word embedding overcomes the limitations of context-counting approaches – due to their poor scalability to large document collections (Mikolov et al. 2013a) – and provides more effective vector representation of words (Baroni et al. 2014, Mikolov et al. 2013b). Thus, in our study, we adopt word embedding for building our distributional semantic model.

Ye et al. (2016) already exploited word embedding for enhancing information retrieval in software engineering. They run word2vec on a collection of over 12K Java SE 7 documents to represent both natural language words and source code tokens in a distributional semantic model. Their final goal is to bridge the lexical gap between code fragments and natural language description that can be found in tutorials, API documentations, and bug reports. They empirically demonstrate how exploiting word embedding improves over state-of-the-art approaches to bug localization. Furthermore, they demonstrate the benefit of exploiting word embedding for linking API documents to Java questions posted in Stack Overflow.

## 10. Conclusions

We presented Senti4SD, a sentiment polarity classifier for software developers' artifacts. The classifier is trained and tested on a gold standard of over 4K posts mined from Stack Overflow and manually annotated with emotion polarity. The gold standard is publicly available for further studies on emotion awareness in software engineering. We also release the guidelines for an annotation to encourage the community to extend and further validate our dataset by replicating the annotation experiment.

The semantic features of Senti4SD are computed based on a distributional semantic model built exploiting word embedding. We built the DSM by running word2vec on a collection of over 20 million documents from Stack Overflow, thus obtaining word vectors that are representative of developers' communication style. The DSM is released, with the replication kit, for future research on word embedding for text categorization and information retrieval in software engineering.

By combining lexicon-based, keyword-based and semantic features, Senti4SD successfully addresses the problem of the negative bias in off-the-shelf sentiment analysis tools. In particular, we observe a 19% improvement in precision for the negative class and a 25% improvement in recall for the neutral class with respect to the baseline represented by SentiStrength.

As future work, we intend to explore the contribution of additional features to capture further meaningful aspects of language use, such as part-of-speech and the rhetorical structure of sentences. We also intend to fine-tune Senti4SD to recognize content with mixed sentiment. As a further assessment of our approach, we intend to evaluate Senti4SD performance on further crowd-generated content from other social software engineering tools and repositories (e.g., GitHub, issue tracking systems). Besides, we plan to provide further benchmarking with other sentiment analysis tools and lexicons. Finally, we are also working on an extended version of our gold standard that will include emotion labels (e.g., love, anger, sadness, joy), as a first step towards building a classifier to detect specific emotions.

**Acknowledgments**

This work is partially supported by the project 'EmoQuest - Investigating the Role of Emotions in Online Question & Answer Sites', funded by the Italian Ministry of Education, University and Research (MIUR) under the program "Scientific Independence of young Researchers" (SIR). The computational work has been executed on the IT resources made available by two projects, ReCaS and PRISMA, funded by MIUR under the program "PON R&C 2007-2013". We thank Pierpaolo Basile for insightful discussions and helpful comments and the annotators involved in the gold standard building.

**References**

Anderson A, Huttenlocher D, Kleinberg J, Leskovec J (2012) Discovering value from community activity on focused question answering sites: A case study of stack overflow. In: Proceedings of the 18th ACM SIGKDD International Conference on Knowledge Discovery and Data Mining, ACM, New York, NY, USA, KDD'12, pp 850–858, DOI 10.1145/2339530.2339665

Asaduzzaman M, Mashiyat AS, Roy CK, Schneider KA (2013) Answering questions about unanswered questions of stack overflow. In: Proceedings of the 10th Working Conference on Mining Software Repositories, IEEE Press, Piscataway, NJ, USA, MSR '13, pp 97–100

Baroni M, Dinu G, Kruszewski G (2014) Don't count, predict! a systematic comparison of context-counting vs. context-predicting semantic vectors. In: Proceedings of the 52nd Annual Meeting of the Association for Computational Linguistics (Volume 1: Long Papers), Association for Computational Linguistics, Baltimore, Maryland, pp 238–247

Barua A, Thomas SW, Hassan AE (2014) What are developers talking about? an analysis of topics and trends in stack overflow. Empirical Softw Engg 19(3):619–654, DOI 10.1007/s10664-012-9231-y
Basile P, Novielli N (2015) Uniba: Sentiment analysis of English tweets combining micro-blogging, lexicon and semantic features. In: Proceedings of the 9th International Workshop on Semantic Evaluation (SemEval 2015), ACL, pp 595–600
Bengio Y, Ducharme R, Vincent P, Janvin C (2003) A neural probabilistic language model. J Mach Learn Res 3:1137–1155
Blaz CCA, Becker K (2016) Sentiment analysis in tickets for IT support. In: Proceedings of the 13th International Conference on Mining Software Repositories, ACM, New York, NY, USA, MSR '16, pp 235–246, DOI 10.1145/2901739.2901781
Calefato F, Lanubile F, Marasciulo MC, Novielli N (2015) Mining successful answers in stack overflow. In: Proceedings of the 12th Working Conference on Mining Software Repositories, IEEE Press, Piscataway, NJ, USA, MSR '15, pp 430–433
Carofiglio V, de Rosis F, Novielli N (2009) Cognitive Emotion Modeling in Natural Language Communication, Springer London, London, pp 23–44
Cohen J (1968) Weighted kappa: Nominal scale agreement provision for scaled disagreement or partial credit. Psychological Bulletin
Collobert R, Weston J (2008) A unified architecture for natural language processing: Deep neural networks with multitask learning. In: Proceedings of the 25th International Conference on Machine Learning, ACM, New York, NY, USA, ICML '08, pp 160–167, DOI 10.1145/1390156.1390177
Danescu-Niculescu-Mizil C, Sudhof M, Jurafsky D, Leskovec J, Potts C (2013) A computational approach to politeness with application to social factors. In: ACL (1), The Association for Computer Linguistics, pp 250–259
Ekman P (1999) Handbook of Cognition and Emotion. John Wiley & Sons Ltd
De Lucia A, Fasano F, Oliveto R, Tortora G (2007) Recovering traceability links in software artifact management systems using information retrieval methods. ACM Trans Softw Eng Methodol 16(4), DOI 10.1145/1276933.1276934
Denning PJ. Moods. (2012) Commun. ACM, 55(12):33–35.
Fan RE, Chang KW, Hsieh CJ, Wang XR, Lin CJ (2008) Liblinear: A library for large linear classification. J Mach Learn Res 9:1871–1874, URL http://dl.acm.org/citation.cfm?id=1390681.1442794
Ford D and Parnin C (2015) Exploring causes of frustration for software developers. In CHASE, pages 115–116. IEEE Press.
Gachechiladze D, Lanubile F, Novielli N, and Serebrenik A (2017). Anger and its direction in collaborative software development. In Proceedings of the 39th International Conference on Software Engineering: New Ideas and Emerging Results Track (ICSE-NIER '17). IEEE Press, Piscataway, NJ, USA, 11-14. DOI: https://doi.org/10.1109/ICSE-NIER.2017.18
Graziotin D, Fagerholm F, Wang X, Abrahamsson P (2017) Unhappy Developers: Bad for Themselves, Bad for Process, and Bad for Software Product. To appear as a poster paper in the Proceedings of the 39th International Conference on Software Engineering (ICSE '17).
Guzman E, Bruegge B (2013) Towards emotional awareness in software development teams. In: Proceedings of the 2013 9th Joint Meeting on Foundations of Software Engineering, ACM, New York, NY, USA, ESEC/FSE 2013, pp 671–674, DOI 10.1145/2491411.2494578
Guzman E, Azocar D, Li Y (2014) Sentiment analysis of commit comments in Github: An empirical study. In: Proceedings of the 11th Working Conference on Mining Software Repositories, ACM, New York, NY, USA, MSR 2014, pp 352–355, DOI 10.1145/2597073.2597118
Guzman E, Alkadhi R, Seyff N (2016) A needle in a haystack: What do twitter users say about software? In: 24th IEEE International Requirements Engineering Conference In: Proceedings of the IEEE 24th International Requirements Engineering Conference (RE), pp. 96-105, doi: 10.1109/RE.2016.67
He H, Garcia EA, Learning from Imbalanced Data. IEEE Transasction on Knowledge and Data Engineering 21(9), 1263–1284 (2009). doi:10.1109/TKDE.2008.239.
Helleputte T (2015) LiblineaR: Linear Predictive Models Based on the LIBLINEAR C/C++ Library. R package version 1.94-2
Hogenboom A, Frasincar F, de Jong F, Kaymak U (2015) Using rhetorical structure in sentiment analysis. Commun ACM 58(7):69–77, DOI 10.1145/2699418
Islam MDR and Zibran MF (2017) Leveraging automated sentiment analysis in software engineering. In Proceedings of the 14th International Conference on Mining Software Repositories (MSR '17). IEEE Press, Piscataway, NJ, USA, 203-214. DOI: https://doi.org/10.1109/MSR.2017.9
Joachims T (1998) Text categorization with suport vector machines: Learning with many relevant features. In: Proceedings of the 10th European Conference on Machine Learning, Springer-Verlag, London, UK, UK, ECML '98, pp 137–142
Joachims T (2006) Training linear SVMs in linear time. In: Proceedings of the 12th ACM SIGKDD International Conference on Knowledge Discovery and Data Mining, ACM, New York, NY, USA, KDD '06, pp 217–226, DOI 10.1145/1150402.1150429
Jongeling R, Datta S, Serebrenik A (2015) Choosing your weapons: On sentiment analysis tools for software engineering research. In: Software Maintenance and Evolution (ICSME), 2015 IEEE International Conference on, pp 531–535, DOI

10.1109/ICSM.2015.7332508

Kucuktunc O, Cambazoglu BB, Weber I, Ferhatosmanoglu H (2012) A large- scale sentiment analysis for Yahoo! answers. In: Proceedings of the Fifth ACM International Conference on Web Search and Data Mining, ACM, New York, NY, USA, WSDM '12, pp 633–642, DOI 10.1145/2124295.2124371

Kuhn M (2016) Contributions from Jed Wing, S. Weston, A. Williams, C. Keefer, A. Engelhardt, T. Cooper, Z. Mayer, B. Kenkel, the R Core Team, M. Benesty, R. Lescarbeau, A. Ziem, L. Scrucca, Y. Tang, and C. Candan., caret: Classification and Regression Training, 2016, r package version 6.0-70. Available: https://CRAN.R- project.org/package=caret

Landauer TK, Dutnais ST (1997) A solution to Platos problem: The latent semantic analysis theory of acquisition, induction, and representation of knowledge. PSYCHOLOGICAL REVIEW 104(2):211–240

Lazarus R (1991) Emotion and adaptation. New York: Oxford University Press.

Levy O, Goldberg Y (2014) Neural word embedding as implicit matrix factorization. In: Ghahramani Z, Welling M, Cortes C, Lawrence ND, Weinberger KQ (Eds) Advances in Neural Information Processing Systems 27, Curran Associates, Inc., pp 2177–2185, URL http://papers.nips.cc/paper/5477-neural-word-embedding-as- implicit-matrix-factorization.pdf

Maalej W, Kurtanovic Z, Nabil H, Stanik C (2016) On the automatic classification of app reviews. Requirements Engineering 21(3):311–331, DOI 10.1007/s00766-016-0251-9

Manning CD, Surdeanu M, Bauer J, Finkel J, Bethard SJ, McClosky D (2014) The Stanford CoreNLP natural language processing toolkit. In: Proceedings of 52nd Annual Meeting of the Association for Computational Linguistics: System Demonstrations, pp 55–60

Mäntylä M, Adams B, Destefanis G, Graziotin D, Ortu M (2016) Mining valence, arousal, and dominance: Possibilities for detecting burnout and productivity? In: Proceedings of the 13th International Conference on Mining Software Repositories, ACM, New York, NY, USA, MSR '16, pp 247–258, DOI 10.1145/2901739.2901752

Mäntylä MV, Novielli N, Lanubile F, Claes M, and Kuutila M (2017) Bootstrapping a lexicon for emotional arousal in software engineering. In Proceedings of the 14th International Conference on Mining Software Repositories (MSR '17). IEEE Press, Piscataway, NJ, USA, 198-202. DOI: https://doi.org/10.1109/MSR.2017.47

Meta (2017). Meta Stack exchange is too harsh to new users. http://meta.stackexchange.com/questions/179003/stack-exchange-is-too-harsh-to- new-users-please-help-them-improve- low-quality-po, Last accessed: February 2017

Mikolov T, Chen K, Corrado G, Dean J (2013a) Efficient estimation of word representations in vector space. CoRR abs/1301.3781

Mikolov T, Sutskever I, Chen K, Corrado GS, Dean J (2013b) Distributed representations of words and phrases and their compositionality. In: Burges CJC, Bottou L, Welling M, Ghahramani Z, Weinberger KQ (Eds) Advances in Neural Information Processing Systems 26, Cur- ran Associates, Inc., pp 3111–3119

Miller GA, Charles WG (1991) Contextual Correlates of Semantic Similarity. Language and Cognitive Processes 6(1):1–28, DOI 10.1080/01690969108406936

Mitchell TM (1997) Machine Learning (1 ed.). McGraw-Hill, Inc., New York, NY, USA.

Mohammad SM (2016) Sentiment analysis: Detecting valence, emotions, and other affectual states from text. In: Meiselman H (Ed) Emotion Measurement, Elsevier

Mohammad SM, Kiritchenko S, Zhu X (2013) NRC-Canada: Building the state-of-the-art in sentiment analysis of tweets. CoRR abs/1308.6242, URL http://arxiv.org/abs/1308.6242

Müller SC and Fritz T (2015) Stuck and frustrated or in flow and happy: sensing developers' emotions and progress. In Proceedings of the 37th International Conference on Software Engineering - Volume 1 (ICSE '15), Vol. 1. IEEE Press, Piscataway, NJ, USA, 688-699.

Murgia A, Tourani P, Adams B, Ortu M (2014) Do developers feel emotions? An exploratory analysis of emotions in software artifacts. In: Proceedings of the 11th Working Conference on Mining Software Repositories, ACM, New York, NY, USA, MSR 2014, pp 262–271, DOI 10.1145/2597073.2597086

Novielli N, Strapparava C (2013) The role of affect analysis in dialogue act identification. IEEE Transactions on Affective Computing 4(4):439– 451, DOI 10.1109/T-AFFC.2013.20

Novielli N, Calefato F, Lanubile F (2014) Towards discovering the role of emotions in Stack Overflow. In Proceedings of the 6th International Workshop on Social Software Engineering (SSE 2014). ACM, New York, NY, USA, 33-36. DOI=http://dx.doi.org/10.1145/2661685.2661689

Novielli N, Calefato F, Lanubile F (2015) The challenges of sentiment detection in the social programmer ecosystem. In: Proceedings of the 7th International Workshop on Social Software Engineering, ACM, New York, NY, USA, SSE 2015, pp 33–40, DOI 10.1145/2804381.2804387

Ortu M, Adams B, Destefanis G, Tourani P, Marchesi M, Tonelli R (2015) Are bullies more productive?: Empirical study of affectiveness vs. issue fixing time. In: Proceedings of the 12th Working Conference on Mining Software Repositories, IEEE Press, Piscataway, NJ, USA, MSR '15, pp 303–313

Ortu M, Murgia A, Destefanis G, Tourani P, Tonelli R, Marchesi M, Adams B (2016) The emotional side of software developers in Jira. In: Proceedings of the 13th International Conference on Mining Software Repositories, ACM, New

York, NY, USA, MSR '16, pp 480–483, DOI 10.1145/2901739.2903505

Pang B, Lee L (2008) Opinion mining and sentiment anal- ysis. Found Trends Inf Retr 2(1-2):1–135, DOI 10.1561/1500000011

Panichella S, Sorbo AD, Guzman E, Visaggio A, Canfora G, Gall H (2015) How can i improve my app? classifying user reviews for software maintenance and evolution. 31st IEEE International Conference on Software Maintenance and Evolution

Pennebaker J and Francis M, Linguistic Inquiry and Word Count: LIWC. Erlbaum Publishers, 2001.

Pletea D, Vasilescu B, and Serebrenik A (2014) Security and emotion: sentiment analysis of security discussions on GitHub. In Proceedings of the 11th Working Conference on Mining Software Repositories (MSR 2014). ACM, New York, NY, USA, 348-351. DOI: http://dx.doi.org/10.1145/2597073.2597117

R Development Core Team (2008) R: A Language and Environment for Statistical Computing. R Foundation for Statistical Computing, Vienna, Austria, URL http://www.R-project.org, ISBN 3-900051-07-0

Rahman MM, Roy CK, Keivanloo I (2015) Recommending insightful comments for source code using crowdsourced knowledge. In: 15th IEEE International Working Conference on Source Code Analysis and Manipulation, SCAM 2015, Bremen, Germany, September 27-28, 2015, pp 81–90, DOI 10.1109/SCAM.2015.7335404

Russell J (1980) A circumplex model of affect. Journal of personality and social psychology 39(6):1161–1178

Saif H, Fernandez M, He Y, Alani H (2014) On stopwords, filtering and data sparsity for sentiment analysis of twitter. In: Chair) NCC, Choukri K, Declerck T, Loftsson H, Maegaard B, Mariani J, Moreno A, Odijk J, Piperidis S (eds) Proceedings of the Ninth International Conference on Language Re- sources and Evaluation (LREC'14), European Language Resources Association (ELRA), Reykjavik, Iceland

Scherer K, Wranik T, Sangsue J, Tran V, Scherer U (2004) Emotions in everyday life: Probability of oc- currence, risk factors, appraisal and reaction patterns. Social Science Information 43(4):499–570

Sebastiani F (2002) Machine learning in automated text categorization. ACM Comput Surv 34(1):1–47, DOI 10.1145/505282.505283

SEmotion (2016) Proceedings of the 1st International Workshop on Emotion Awareness in Software Engineering, ACM, New York, NY, USA

Shaver P, Schwartz J, Kirson D, O'Connor C (1987) Emotion knowledge: Further exploration of a prototype approach. Journal of Personality and Social Psychology 52(6):1061–1086, DOI 10.1037//0022- 3514.52.6.1061, URL http://dx.doi.org/10.1037//0022-3514.52.6.1061

Sinha V, Lazar A, Sharif B (2016) Analyzing developer sentiment in commit logs. In: Proceedings of the 13th International Conference on Mining Software Repositories, ACM, New York, NY, USA, MSR '16, pp 520–523, DOI 10.1145/2901739.2903501

Smolensky P (1990) Tensor product variable binding and the representation of symbolic structures in connectionist systems. Artificial Intelligence 46(1-2):159–216, DOI 10.1016/0004-3702(90)90007-M

Socher R, Perelygin A, Wu J, Chuang J, Manning CD, Ng AY, Potts C (2013) Recursive deep models for semantic compositionality over a sentiment treebank. In: Proceedings of the 2013 Conference on Empirical Methods in Natural Language Processing, Association for Computational Linguistics, Stroudsburg, PA, pp 1631–1642

Strapparava C, Valitutti A (2004) WordNet-Affect: an affective extension of WordNet. In: Proceedings of LREC, vol 4, pp 1083–1086

Stone PJ, Dunphy DC, Smith MS, and Ogilvie DM (1966). The general inquirer: A computer approach to content analysis. Cambridge, MA: The MIT Press.

Thelwall M, Buckley K, Paltoglou G (2012) Sentiment strength detection for the social web. J Am Soc Inf Sci Technol 63(1):163–173, DOI 10.1002/asi.21662

Tian Y, Lo D, Lawall J (2014) Sewordsim: Software-specific word similarity database. In: Companion Proceedings of the 36th International Conference on Software Engineering, ACM, New York, NY, USA, ICSE Companion 2014, pp 568–571, DOI 10.1145/2591062.2591071

Tromp E and Pechenizkiy M (2015) Pattern-based emotion classification on social media. In M. M. Gaber, M. Cocea, N. Wiratunga, and A. Goker, editors, *Advances in Social Media Analysis*, pages 1–20. Springer.

Wittgenstein L (1965) Philosophical Investigations. The Macmillan Company, New York, NY, USA

Ye X, Shen H, Ma X, Bunescu RC, Liu C (2016) From word embeddings to document similarities for improved information retrieval in software engineering. In: Proceedings of the 38th International Conference on Software Engineering, ICSE 2016, Austin, TX, USA, May 14-22, 2016, pp 404–415, DOI 10.1145/2884781.2884862

**Appendix A: Coding Guidelines**

In the following, we report the task description and the guidelines used for training the coders involved in the emotion annotation study.

*Task Description and Annotation Guidelines.* You are invited to take part in the annotation study of developers contributed texts in Stack Overflow. We are interested in annotating the presence of emotions in technical documents authored by developers during their online interactions.

The data source is the official Stack Overflow dump released by Stack Exchange on May '15. You will be required to annotate randomly selected posts, including questions, answers, and comments. The unit of annotation is the entire post.

You will use the coding schema reported in TABLE A. For each post, please indicate what emotion it conveys (if any) among the basic emotions (first column in the table), that are, *love, joy, surprise, anger, sadness,* and *fear*. Multiple Emotion labels are allowed but you should try to avoid if possible. You can use the second and third level in the schema as a reference for choosing the primary emotion, as shown in TABLE B.

Once you define the emotion label, please specify the emotion polarity accordingly, choosing among *positive, negative, neutral,* and *mixed*. If the post does not contain any emotion, it should be annotated as neutral. The surprise is the only emotion that could match any of the polarity value: please, carefully evaluate each post in order to determine if it conveys positive, negative, or neutral polarity. If multiple emotion labels are indicated in a given text, you should define the polarity accordingly. A text annotated with one or more positive emotions only has a positive polarity. Conversely, a post annotated with one or more negative emotions holds a negative polarity. If both positive and negative emotions are found, you should indicate both. If you wish to indicate a polarity label you are required to specify the corresponding emotion. The absence of emotion can be annotated exclusively as neutral. The list of all possible combination allowed and not allowed by our coding schema is reported in TABLE C.

TABLE A. MAPPING THE SHAVER ET AL. EMOTION FRAMEWORK TO SENTIMENT POLARITY

| Emotion Polarity | Basic Emotions | Second level Emotions | Third level Emotions |
|---|---|---|---|
| Positive | Love | Affection | Liking, Caring, Compassion, Fondness, Affection, Love, Attraction, Tenderness, Sentimentality |
| | | Lust | Desire, Passion, Infatuation, Arousal |
| | | Longing | --- |
| | Joy | Cheerfulness | Happiness, Amusement, Satisfaction, Bliss, Gaiety, Glee, Jolliness, Joviality, Joy, Delight, Enjoyment, Gladness, Jubilation, Elation, Ecstasy, Euphoria |
| | | Zest | Enthusiasm, Excitement, Thrill, Zeal, Exhilaration |
| | | Contentment | Pleasure |
| | | Optimism | Hope, Eagerness |
| | | Pride | Triumph |
| | | Enthrallment | Rapture |
| Negative | Anger | Irritation | Annoyance, Agitation, Grumpiness, Aggravation, Grouchiness |
| | | Exasperation | Frustration |
| | | Rage | Anger, Fury, Hate, Dislike, Resentment, Outrage, Wrath, Hostility, Bitterness, Ferocity, Loathing, Scorn, Spite, Vengefulness |
| | | Disgust | Revulsion, Contempt |
| | | Envy | Jealousy |
| | | Torment | --- |
| | Sadness | Suffering | Hurt, Anguish, Agony |
| | | Sadness | Depression, Sorrow, Despair, Gloom, Hopelessness, Glumness, Unhappiness, Grief, Woe, Misery, Melancholy |
| | | Disappointment | Displeasure, Dismay |
| | | Shame | Guilt, Regret, Remorse |
| | | Neglect | Embarrassment, Insecurity, Insult, Rejection, Alienation, Isolation, Loneliness, Homesickness, Defeat, Dejection, Humiliation |
| | | Sympathy | Pity |
| | Fear | Horror | Alarm, Fright, Panic, Terror, Fear, Hysteria, Shock, Mortification |
| | | Nervousness | Anxiety, Distress, Worry, Uneasiness, Tenseness, Apprehension, Dread |
| Either Positive or Negative | Surprise | Surprise | Amazement, Astonishment |

TABLE B. EXAMPLES OF ANNOTATED POSTS.

| Input Text | Annotation | | Rationale for annotation *(second and third level emotion found)* |
|---|---|---|---|
| | *Basic Emotion(s) found* | *Polarity* | |
| *"Thanks for your input! You're, like, awesome"* | Love | Positive | Liking (third level), Affection (second level) indicating gratitude. |
| *"I'm happy with the approach and the code looks good"*. | Joy | Positive | Happiness, Satisfaction (third), Cheerfulness (second) |
| *"I still question the default, which can lead to surprisingly huge memory use"* | Surprise | Negative | Surprise (second) due to the unexpected undesirable behavior of the code. |
| *"I will come over to your work and slap you "* | Anger | Negative | Hostility (third), Rage (second) |
| *"Sorry for the delay Stephen"* | Sadness | Negative | Guilt (third), Shame (second) |
| *"I'm worried about some subtle differences between char and Character"* | Fear | Negative | Worry (third) |

TABLE C. COMBINATIONS OF VALUES ALLOWED AND NOT ALLOWED BY OUR ANNOTATION SCHEMA.

| Love | Joy | Surprise | Anger | Sadness | Fear | Polarity | Explanation |
|---|---|---|---|---|---|---|---|
| **Annotation allowed by our schema** | | | | | | | |
| | | | x | | | Negative | *Negative emotion and negative polarity* |
| | x | | | | | Positive | *Positive emotion and positive polarity* |
| | | x | | | | Negative | |
| | | x | | | | Positive | *Surprise is intrinsically ambiguous, all polarity values are allowed* |
| | | x | | | | Neutral | |
| x | x | | | | | Positive | *Multiple emotion labels, positive polarity* |
| | | | x | x | | Negative | *Multiple emotion labels, negative polarity* |
| | x | | x | | | Mixed | *Multiple emotion labels, mixed polarity* |
| | | | | | | Neutral | *Absence of emotion* |
| **Annotation NOT allowed by our schema** | | | | | | | |
| | | | | | | Negative | *No emotion and negative polarity* |
| | | | | | | Positive | *No emotion and positive polarity* |
| | | | | | | Mixed | *No emotion and mixed polarity* |
| x | | | | | | Neutral | *Emotion label different from surprise and neutral polarity* |
| | | | | x | | Neutral | |